\documentclass[prb,twocolumn,showpacs]{revtex4}
\usepackage{amsmath}
\usepackage{amssymb}
\usepackage{dcolumn}
\usepackage{graphicx}

\begin{document}

\title[Short title for running header]{Theory of the giant thermal Hall effect in the high temperature superconductors}
\author{Tao Li}
\affiliation{Department of Physics, Renmin
University of China, Beijing 100872, P.R.China}
\date{\today}

\begin{abstract}
Recent transport measurement finds giant negative thermal Hall signal in the pseudo-gap phase of the high temperature superconductors\cite{Taillefer}. Such a signal is found to increase in magnitude with decreasing doping and to reach its maximum in the half-filled antiferromagnetic ordered parent compounds. It is still elusive what is the implication of such a phenomena for the mechanism of the pseudo-gap phase. The observation of such a signal may also challenge our well established understanding of the parent compounds as perfect Heisenberg antiferromagnets. Here we show that such a giant thermal Hall effect can be naturally understood as the orbital magnetic response of a quantum Heisenberg antiferromagnet with sizable multi-spin exchange coupling. We also argue that the observation of the giant thermal Hall effect in the pseudo-gap phase imply that the origin of the pseudo-gap in the high temperature superconductors is closely related to the antiferromagnetic correlation between local spins.    
\end{abstract}

\maketitle
The origin of the pseudo-gap phase in the high temperature superconductors is still under intense debate. There are many clues indicating the importance of antiferromagnetic correlation in the development of the pseudo-gap phenomena. For example, ARPES measurement find that the pseudo-gap in the anti-nodal region opens right at the temperature when the $^{63}Cu$ NMR spin relaxation rate reaches its maximum\cite{Hashimoto}. More recently, thermodynamic measurement find that the pseudo-gap phase end suddenly at a quantum critical point that is far from any known ordering instability\cite{Michon}. This observation is, however, consistent with the picture of a system with antiferromagnetic short-range correlated local spin coupled to an itinerant quasiparticle system at the special doping when the Van Hove singularity meet the antiferromagnetic hot spot\cite{Li}.    

Very recently, transport measurement on various high temperature superconductors find unexpected giant thermal Hall sign just below the critical doping for pseudo-gap phase\cite{Taillefer}. The sign of the thermal Hall signal is found to be opposite to that of the charge carriers and the size of the thermal Hall signal is found to increase with decreasing doping and to reach its maximum at half filling. These observations indicate that the signal come from degree of freedom other than the itinerant quasiparticles. The experiment suggests that the giant thermal Hall effect is a generic property of the pseudo-gap phase of the high temperature superconductors. A very natural conclusion is that the giant thermal Hall effect should be contributed by the local spin degree of freedom. importantly, the the observed thermal Hall signal is found to increase linearly with the applied magnetic field. The signal also increases monotonically with decreasing temperature.

The local spin degree of freedom in the parent compounds of the high temperature superconductors is known to be well described by the Heisenberg antiferromagnetic model. These local spin degree of freedom and their antiferromagnetic local correlation remains robust against fairly large doping, as implied by the magnon-like dispersive mode revealed by recent RIXS measurements\cite{RIXS}. However, it is unclear why such seemingly simple Heisenberg antiferromagnet can contribute such a giant thermal Hall effect. Previous theoretical study of the thermal Hall effect in quantum magnet suggest that on the square lattice most mechanisms proposed for the thermal Hall effect are ineffective as a result of cancellation of chiral contributions from edge-sharing plaquettes\cite{PALee2,Murakami,Hotta,PALee,Sachdev,Xu,ZXLi}. For example, the effect of both the DM interaction and the orbital magnetic coupling are argued to give zero contribution to the thermal Hall effect for the high temperature superconductors, no matter if a magnon mechanism or a spinon mechanism is assumed.

Here we show that the observed giant thermal Hall effect can indeed be accounted for by the orbital magnetic response of a quantum antiferromagnet on the square lattice. Here we single out the orbital magnetic coupling but neglect the combined effect of the Zeeman coupling and the anisotropic DM interaction. This is reasonable as has been shown earlier that with the spin rotational symmetry, time reversal symmetry and the parity symmetry, the combined effect of the  Zeeman coupling and the DM interaction can not account for a sizable thermal Hall effect for the realistic DM coupling in the high temperature superconductors.    

We start from the spin-1/2 Heisenberg model on the square lattice with the Hamiltonian
\begin{equation}
H_{J}=J\sum_{<i,j>} \mathrm{S}_{i} \cdot \mathrm{S}_{j}.
\end{equation}
Here the sum is over nearest neighboring sites. We define $|\mathrm{HAF}\rangle$ as the ground state of $H_{J}$ for latter convenience. Further neighboring exchange couplings are assumed to be negligible. In the high temperature superconductors, the next-nearest-neighboring electron hopping term $t'$ is about $1/4$ of the nearest-neighbor hopping term $t$. We thus expect the next-nearest-neighboring exchange coupling $J'$  to be about $J/16$. The external magnetic field can couple to the local spin through either the Zeeman term or through the multi-spin exchange process in terms of the spin chirality. As we stated above, we will neglect the Zeeman coupling in this study. Up to the fourth order in the perturbative expansion of the Hubbard model in the large $U$ limit, the coupling of the magnetic field to the spin chirality takes the form of
\begin{equation}
H_{c}=J_{3}\sin(\Phi_{3})\ H_{3}+J_{4}\sin(\Phi_{4})\ H_{4}.
\end{equation} 
Here $J_{3}=-\frac{24t^{2}t'}{U^{2}}$ is the three spin exchange coupling around an elementary triangle $\bigtriangleup_{i,j,k}$ on the square lattice,  $J_{4}=80t^{4}/U^{3}$ is the four spin exchange coupling around an elementary plaquette $\square_{i,j,k,l}$ on the square lattice(here the group of sites $i,j,k$ in the triangle $\bigtriangleup_{i,j,k}$ and $i,j,k,l$ in the palquette $\square_{i,j,k,l}$ are both ordered in the anti-clockwise fashion). $\Phi_{3}$ and $\Phi_{4}$ are the gauge fluxes of the external magnetic field enclosed in $\bigtriangleup_{i,j,k}$ and $\square_{i,j,k,l}$. $H_{3}$ and $H_{4}$ are given by
\begin{eqnarray}
H_{3}&=&\sum_{\bigtriangleup_{i,j,k}} \ \mathrm{S}_{i}\cdot (\mathrm{S}_{j}\times\mathrm{S}_{k})\nonumber\\
H_{4}&=&i\sum_{\square_{i,j,k,l}} (P_{i,j,k,l}-P^{-1}_{i,j,k,l}).
\end{eqnarray}
Here $P_{i,j,k,l}$ denotes the anti-clockwise ring exchange of the four spins on the elementary plaquette $\square_{i,j,k,l}$.  According to the experimental fit for the parent compound La$_{2}$CuO$_{4}$, $J_{4}\approx 40$ meV, which is not at all small\cite{LCO}. We note that both $H_{3}$ and $H_{4}$ share the same symmetry property with the thermal Hall coefficient $\kappa_{xy}$, namely, they are all odd under time reversal and parity operation but invariant under spin rotation.  A nonzero expectation value of either $H_{3}$ or $H_{4}$ in the system is sufficient to generate a thermal Hall response from symmetry point of view.

Previous studies have analyzed the effect of $H_{3}$(or $H_{4}$) on thermal Hall effect in both the magnon picture and the spinon picture\cite{PALee2,Murakami,Hotta,PALee,Sachdev,Xu,ZXLi}. In the magnon picture, it is easy to see that the contributions of $H_{3}$(or $H_{4}$) from different edge-sharing plaquettes cancel with each other on the square lattice at the mean field level.  The situation is different in the spinon picture, in which the cancellation depends on the form the RVB mean field ansatz\cite{Sachdev}. 

Here we will adopt the Bosonic RVB theory\cite{Fermion}, in which the spin operator is written as $\mathrm{S}_{i}=\frac{1}{2}\sum_{\alpha,\beta}b^{\dagger}_{i,\alpha}\sigma_{\alpha,\beta}b_{i,\beta}$. $b_{i,\alpha}$ is a Schwinger Boson operator with $\alpha$ as its spin index. The Schwinger Boson should be subjected to the single occupancy constraint of the form  $\sum_{\alpha}b^{\dagger}_{i,\alpha}b_{i,\alpha}=1$ to be a faithful representation of the spin algebra. With the Schwinger Boson operator, we can define two spin rotationally invariant operators describing the spin correlation between a pair of sites $i$ and $j$. They are the pairing field
\begin{equation}
\hat{A}_{i,j}=\frac{1}{\sqrt{2}}\sum_{\alpha}\ \alpha \ b_{i,\alpha}b_{j,-\alpha}
\end{equation}
describing the antiferromagnetic correlation between site $i$ and $j$ and the hopping field
\begin{equation}
\hat{B}_{i,j}=\frac{1}{\sqrt{2}}\sum_{\alpha}\  \ b^{\dagger}_{i,\alpha}b_{j,\alpha}
\end{equation}
describing the ferromagnetic correlation between site $i$ and $j$. The Heisenberg exchange coupling term $\mathrm{S}_{i}\cdot\mathrm{S}_{j}$ can be either written as 
$-\frac{1}{2}\hat{A}^{\dagger}_{i,j}\hat{A}_{i,j}+C$ or $\frac{1}{2}\hat{B}^{\dagger}_{i,j}\hat{B}_{i,j}+C'$, with $C$ and $C'$ two constants.
For the antiferromagnetic Heisenberg model on the square lattice, the Hamiltonian is most naturally written in the form of
\begin{equation}
H=-\frac{J}{2}\sum_{<i,j>}\hat{A}^{\dagger}_{i,j}\hat{A}_{i,j}+C,
\end{equation}
since we only expect antiferromagnetic correlation between nearest neighboring sites. This form is suggestive of a mean field decoupling in terms of the pairing field $A_{i,j}=<\hat{A}_{i,j}>$. More specifically, the mean field ansatz for this saddle point is given by
\begin{eqnarray}
A_{i,i+x}=A_{i,i+y}=A\nonumber\\
B_{i,i+x}=B_{i,i+y}=0.
\end{eqnarray}
Note that the pairing field $A_{i,j}$ is anti-symmetric and translational invariant in this ansatz. At the same time, we note that the expectation value of the hopping field $\hat{B}_{i,j}$ between sites in the same sub-lattice(for example between next-nearest-neighboring sites) is nonzero, although these expectation values do not enter the mean field Hamiltonian. Such an ansatz has been found to be extremely accurate for the pure Heisenberg model on the square lattice. The key issue to be answered in this study is whether a linear-in-field thermal Hall response can be generated around this saddle point when we turn on $H_{3}$(or $H_{4}$).  

The spin chirality term $H_{3}$ also has multiple representations in terms of $\hat{A}_{i,j}$ and $\hat{B}_{i,j}$. The most commonly used representation of the spin chirality operator is given by\cite{Messio} 
\begin{equation}
\mathrm{S}_{i}\cdot (\mathrm{S}_{j}\times\mathrm{S}_{k})=\frac{i}{\sqrt{2}}(\hat{B}_{i,j}\hat{B}_{j,k}\hat{B}_{k,i}-h.c.).
\end{equation}
Such a representation makes it clear the relation between the spin chirality on the triangle $\bigtriangleup_{i,j,k}$ and the spin Berry phase of a spin-1/2 particle moving around it. One problem with such a representation is that at the mean field level the effect of $H_{3}$ depends quadratically on the expectation value of $\hat{B}_{i,j}$ between nearest neighboring sites, which is zero in the absence of the magnetic field. We thus expect that $H_{3}$ can at most contribute a thermal Hall signal cubic in the strength of the magnetic field. A second representation of $H_{3}$ involving both the pairing field $\hat{A}_{i,j}$ and the hopping field $\hat{B}_{i,j}$ is given by
\begin{eqnarray}
\mathrm{S}_{i}\cdot (\mathrm{S}_{j}\times\mathrm{S}_{k})=\frac{1}{3\sqrt{2}i} \{\hat{A}^{\dagger}_{k,i}\hat{A}_{i,j}\hat{B}_{j,k}
+\hat{A}^{\dagger}_{i,j}\hat{A}_{j,k}\hat{B}_{k,i}\nonumber\\
+\hat{A}^{\dagger}_{j,k}\hat{A}_{k,i}\hat{B}_{i,j}-h.c.\}.
\end{eqnarray}  
In such a representation, it seems promising that $H_{3}$ can generate a linear-in-field thermal Hall signal. More specifically, around the triangle $\bigtriangleup_{i,j,k}$ shown in Fig.1, the expectation value of $\hat{A}_{i,j}, \hat{A}_{j,k}$ and $\hat{B}_{k,i}$ are all nonzero in the ansatz Eq.(7). We can thus modify their phases to accommodate a nonzero gauge flux $\Phi_{i,j,k}=\arg (A^{*}_{i,j}A_{j,k}B_{k,i})$. However, one find that no matter how we adjust the phases of the six fields  $A_{i,j}, A_{j,k}, A_{l,k}, A_{il}, B_{j,k}$ and $B_{i,l}$, we always have $\Phi_{i,j,k}+\Phi_{j,k,l}+\Phi_{k,l,i}+\Phi_{l,i,j}=0\mod(2\pi)$. The coupling of the magnetic field to the spin chirality in $H_{3}$ thus again vanishes at the linear order in the mean field approximation.
\begin{figure}
\includegraphics[width=6cm]{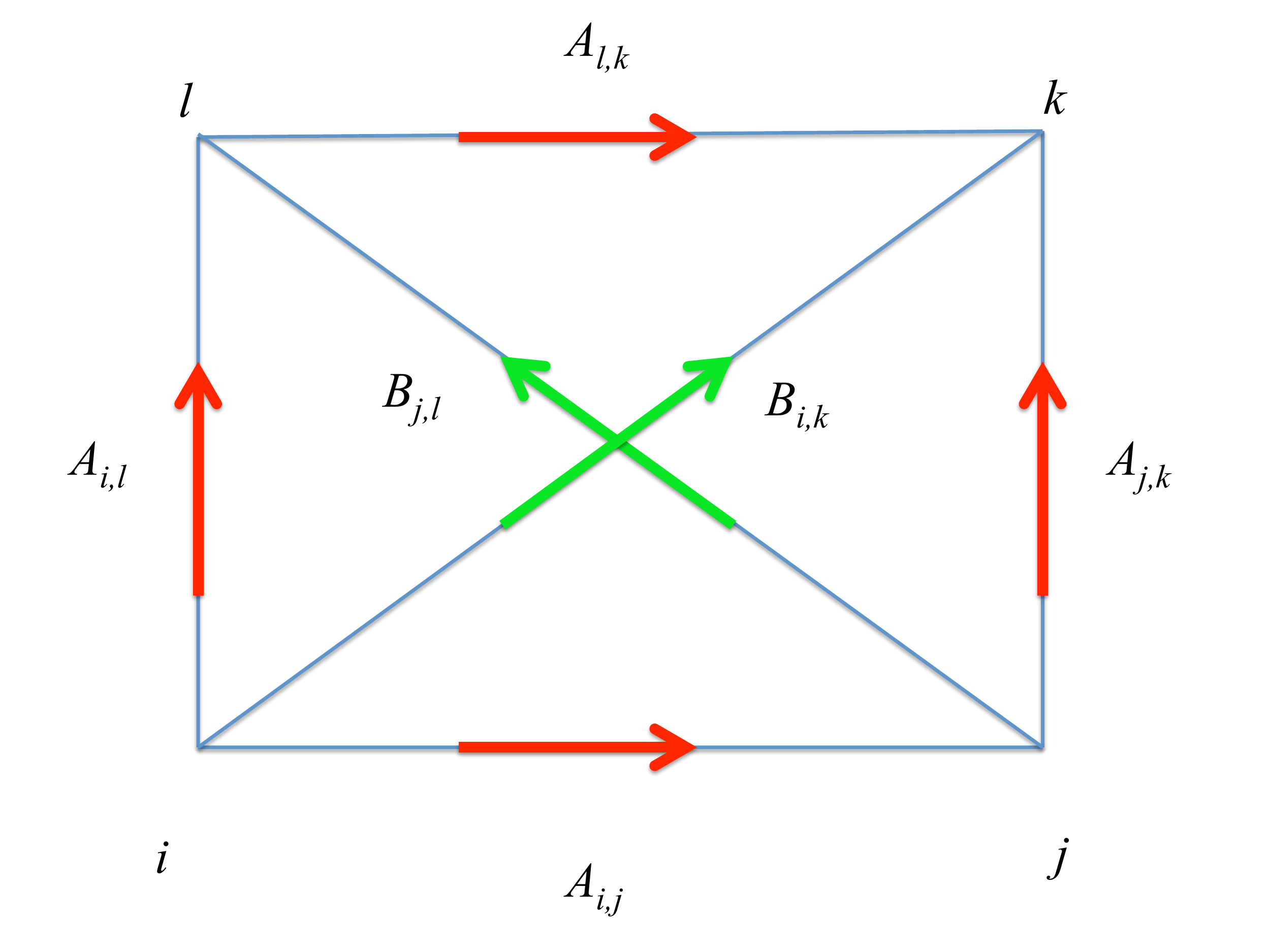}
\caption{\label{fig1}
(Color on-line)Demonstration of the cancellation at the linear order of the mean field contributions from $H_{3}$ on the square lattice. Here we assume the conventional ansatz of the antiferromagnetic Heisenberg model on the square lattice with nonzero $A_{i,j}$ but zero $B_{i,j}$ on nearest neighboring bonds and nonzero $B_{i,j}$ but zero $A_{i,j}$ on next nearest neighboring bonds. The phases of the pairing and hopping fields can be modified by the magnetic field. The gauge flux enclosed in a triangle(say, $\bigtriangleup_{i,j,k}$) is defined as $\Phi_{i,j,k}=\arg(A^{*}_{i,j}A_{j,k}B_{k,i})$. It is then straightforward to show that no matter how we adjust the phases of the pairing and the hopping fields, we always have $\Phi_{i,j,k}+\Phi_{j,k,l}+\Phi_{k,l,i}+\Phi_{l,i,j}=0\mod(2\pi)$.}
\end{figure}
Now let us check the situation for $H_{4}$.  In the Schwinger Boson representation, we have
\begin{equation}
P_{i,j,k,l}-P^{-1}_{i,j,k,l}=4(\hat{B}_{i,j}\hat{B}_{j,k}\hat{B}_{k,l}\hat{B}_{l,i}-h.c.).
\end{equation}
Here $\hat{B}_{i,j}$ denotes the hopping parameter on nearest neighboring bonds. It thus seems even more hopeless for $H_{4}$ to contribute a thermal Hall response linear in the applied magnetic field if we admit the accuracy of the zero field ansatz Eq.(7).

The above analysis indicates that both $H_{3}$ and $H_{4}$ can not account for the observed linear-in-field thermal Hall signal at the Schwinger Boson mean field level. For sake of completeness, we have performed simulated annealing search of Schwinger Boson mean field ansatz on finite cluster of square lattice for $H=H_{J}+H_{c}$. We find that even for rather large $J_{3}\sin\Phi_{3}$(or $J_{4}\sin\Phi_{4}$), the best ansatz is still Eq.(7) and the expectation value of $H_{c}$ is exactly zero in the optimized ansatz. Thus, for small external field $H_{c}$ has no effect at all in the mean field approximation. In other words, there is no hope to find a mean field ansatz that can optimize the ground state energy and accommodate a finite spin chirality simultaneously on the square lattice. In Ref.\onlinecite{Sachdev}, a Schwinger Boson mean field ansatz with a build-in $\pi$-flux structure is found to be compatible with a linear-in-field thermal Hall response if one introduce $H_{c}$. However, it is well known that the uniform ansatz Eq.(7) is the most accurate mean field description of the antiferromagnetic Heisenberg model on the square lattice. For example, the relative error in the variational ground state energy calculated from the corresponding Gutzwiller projected wave function\cite{LDA} is less than $10^{-4}$. On the other hand, the $\pi$-flux ansatz in the Schwinger Boson theory has a very high energy\cite{Li1}.   

However, such a conclusion is obviously an artifact of the mean field treatment. In fact, although the expectation value of $H_{3}$ and $H_{4}$ are both zero in $|\mathrm{HAF}\rangle$, they can not annihilate  $|\mathrm{HAF}\rangle$,  but can generate virtual excitation on it. In such a situation, the expectation value of $H_{3}$(or $H_{4}$) should in general be linear in the magnetic field when we turn on $H_{c}$. On the other hand, a nonzero expectation value of either $H_{3}$ or $H_{4}$ is all what we need to generate a thermal Hall response from the symmetry point of view. We thus expect trivially a linear-in-field thermal Hall signal in the presence of the orbital coupling $H_{c}$. When the spectrum of $H_{J}$ is gapless, as is indeed the case in the thermodynamic limit, we can even expect a divergent thermal Hall signal in the zero temperature limit. Such a divergence will be cut off in real material for reasons not considered in this study, for example, the Zeeman coupling and exchange anisotropies.

More specifically, the expectation value of the spin chirality operator(for example, $H_{4}$) can be estimated from perturbation theory for small external field as 
\begin{equation}
\langle H_{4} \rangle\simeq J_{4}\sin{\Phi_{4}}\sum_{n\neq0}\frac{|\langle n |H_{4}|\mathrm{HAF}\rangle|^{2}}{E_{0}-E_{n}}.
\end{equation}
Here $| n \rangle$ denotes excited states of the pure Heisenberg antiferromagnetic model on the square lattice. We thus have the following upper limit for the induced spin chirality in the external field
\begin{equation}
\langle H_{4} \rangle \leq J_{4}\sin{\Phi_{4}} \langle H^{2}_{4}\rangle_{\mathrm{HAF}}/\Delta.
\end{equation}
Here $\langle H^{2}_{4}\rangle_{\mathrm{HAF}}$ measures the strength of the fluctuation of $H_{4}$ in $|\mathrm{HAF}\rangle$, $\Delta$ is the gap of the system. At finite temperature, such a gap can be roughly understood as a thermal disordering gap in the Schwinger Boson mean field theory. The gap can also be induced by finite size effect or spin anisotropies.

We thus conclude that the linear-in-field thermal Hall effect in the cuprates should be understood as a purely quantum effect, which is beyond both the semiclassical magnon picture and the Schwinger Boson mean field theory description. More specifically, while the average of the spin chirality is zero in the ground state of the Heisenberg antiferromagnetic model on the square lattice, the quantum fluctuation in the spin chirality is nonzero. A small external field can thus polarize the ground state and induce a linear-in-field spin chirality.  Neither the semiclassical approximation nor the Schwinger Boson mean field theory can describe such a polarization effect self-consistently. 

To substantiate such an understanding, we have performed exact diagonalization calculation on the model $H_{J}+J_{4}\sin(\Phi_{4})H_{4}$ on a $4\times4$ cluster of square lattice. The expectation value of $H_{4}$ is shown in Fig.2 as a function of $J_{4}\sin(\Phi_{4})$. A linear-in-field behavior is clearly demonstrated. The $4\times4$ cluster has a rather large finite size gap. For larger clusters, we expect even larger initial slope of $\langle H_{4}\rangle$ as a function of $J_{4}\sin(\Phi_{4})$.    
\begin{figure}
\includegraphics[width=8cm]{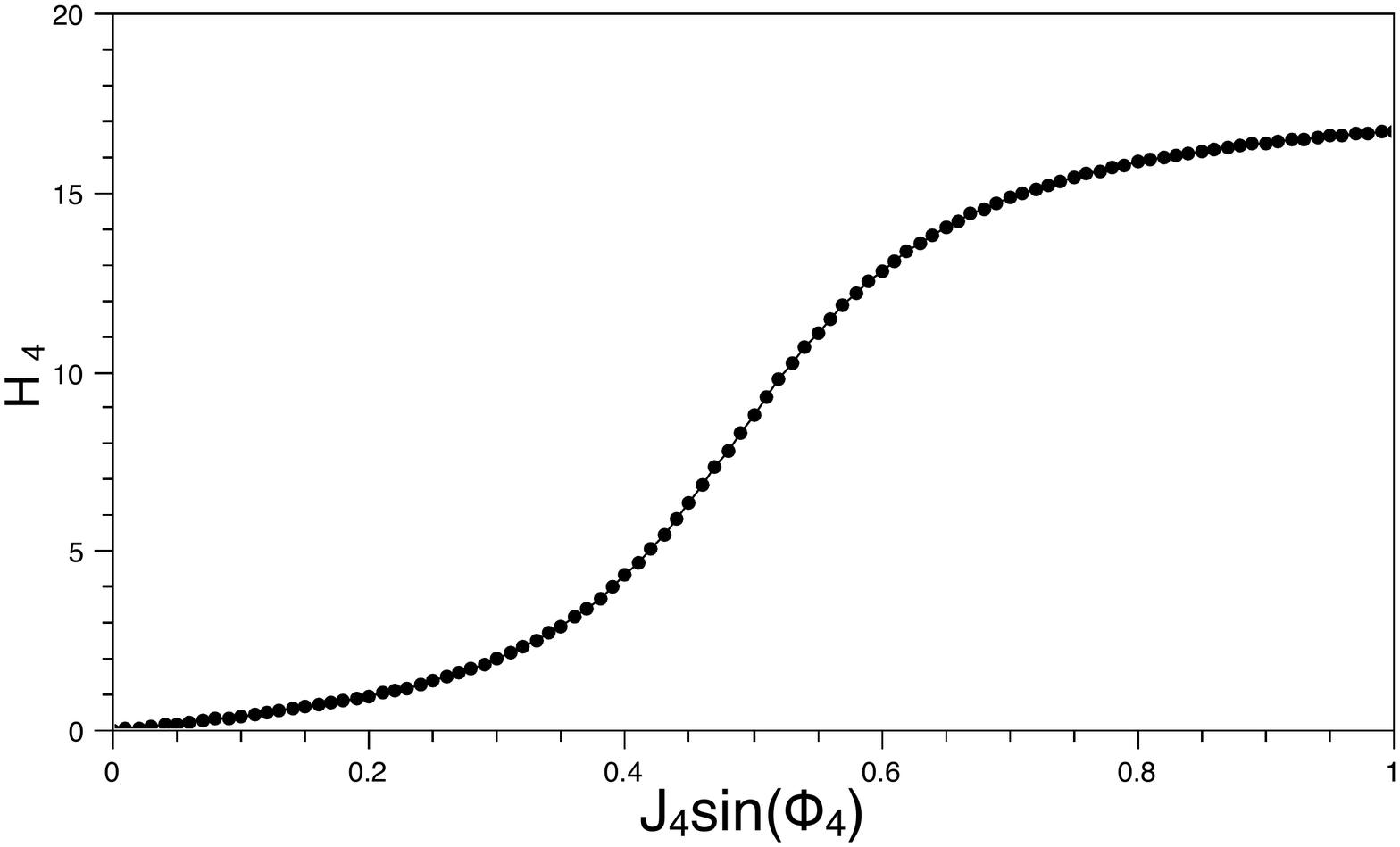}
\caption{\label{fig2}
The linear-in-field spin chirality for small external field as calculated from the exact diagonalization of the model $H_{J}+J_{4}\sin(\Phi_{4})H_{4}$ on a $4\times4$ cluster. Here we have set $J=1$ as the unit of energy.}
\end{figure}

Now we discuss the sign and magnitude of $\kappa_{xy}$ induced by $H_{c}$. Since $t'/t<0$ in the cuprates, both $H_{3}$ and $H_{4}$ contribute to the thermal Hall signal with the same sign. On the other hand,  as $H_{4}$ just describes the collective hopping of four electrons on an elementary plaquette, we expect the thermal Hall current to be deflected in the same fashion as a free electron would be. This explains why we observe a negative $\kappa_{xy}$ in the cuprates. The magnitude of $\kappa_{xy}$ depends on the value of both $J_{3}$ and $J_{4}$. As we mentioned above, such couplings are quite large in the cuprates. This explains why we expect a large $\kappa_{xy}$ in the cuprates. In Ref.\onlinecite{PALee}, it is argued that the effect of $H_{3}$ and $H_{4}$ are both too small at the experimental achievable magnetic field to explain the observed magnitude of the thermal Hall signal, which is estimated to be about $\kappa_{xy}/T\simeq k^{2}_{B}/\hbar$ per CuO$_{2}$ sheet at low temperature. However, we note that such an argument is valid only when we have a large spectral gap(or a much reduced matrix element for the operator $H_{3,4}$). Indeed, one find that the thermal Hall signal in the overdoped regime under the same magnetic field has the same order of magnitude as that observed in the undoped parent compound. The thermal Hall signal in the overdoped system can clearly be attributed to the itinerant charge carrier in a metallic phase, since it follows perfectly the Wiedemann-Franz law. Thus to understand thermal Hall response of the parent compound, it would be helpful to think the system(the Heisenberg antiferromagnet on the square lattice) as a spin analog of a metal, rather than a charge insulator. 

From the above discussion, we find that a linear-in-field thermal Hall signal is naturally expected for a quantum antiferomagnet on the square lattice with a sizable multi-spin exchange coupling. The persistence of the negative thermal Hall signal in the whole pseudo-gap phase of the cuprates then imply strongly the close relationship between the antiferromagnetic correlation and the origin of the pseudo-gap phenomena. Such an understanding is in accordance with a picture for the pseudo-gap end point advocated by us recently\cite{Li}, in which the critical behavior at the pseudo-gap end point is understood as a result of the singular coupling effect between a non-critical local moment system and a quasiparticles system with coincident antiferromagnetic hot spot and Van Hove singularity. 

We acknowledge the support from the grant NSFC 11674391 and the Research Funds of Renmin University of China. The author would also like to thank Masataka Kawano and Chisa Hotta for helpful discussions. We are also grateful to DungHai Lee, Patrick Lee and Subir Sachdev for their invaluable comments.


\begin{thebibliography}{99}
\bibitem{Taillefer} G. Grissonnanche et al., arXiv:1901.03104 (2019).
\bibitem{Hashimoto}M. Hashimoto, R.-H. He, K. Tanaka, J.-P. Testaud, W. Meevasana, R. G. Moore, D.H. Lu, H. Yao, Y. Yoshida, H. Eisaki, T. P. Devereaux, Z. Hussain and Z.-X. Shen, Nat. Phys. \textbf{6}, 414 (2010).
\bibitem{Michon}B. Michon, C. Girod, S. Badoux, J. Kamark, Q. Ma, M. Dragomir, H. A. Dabkowska, B. D. Gaulin, J.-S. Zhou, S. Pyon, T. Takayama, H. Takagi, S. Verret, N.
Doiron-Leyraud, C. Marcenat, L. Taillefer and T. Klein, arXiv:1804.08502, Nature \textbf{567},218 (2019).
\bibitem{Li}T. Li , arXiv:1805.06395 (2018).
\bibitem{RIXS}M. Le Tacon, G. Ghiringhelli, J. Chaloupka, M. M. Sala, V. Hinkov, M. W. Haverkort, M. Minola, M. Bakr, K. J. Zhou, S. Blanco-Canosa, C. Monney, Y. T. Song, G. L. Sun, C. T. Lin, G. M. De Luca, M. Salluzzo, G. Khaliullin, T. Schmitt, L. Braicovich, and B. Keimer, Nat. Phys. \textbf{7}, 725 (2011).
\bibitem{PALee2}H. Katsura, N. Nagaosa, and P. A. Lee,  Phys. Rev. Lett. \textbf{104}, 066403 (2010).
\bibitem{Murakami}R. Matsumoto, R. Shindou, and S. Murakami, Phys. Rev. B \textbf{89}, 054420 (2014).
\bibitem{Hotta}M. Kawano and C. Hotta, Phys. Rev. B \textbf{99}, 054422 (2019).
\bibitem{PALee} J. H. Han, J. -H. Park, and P. A. Lee, arXiv: 1903.01125 (2019).
\bibitem{Sachdev}R. Samajdar, S. Chatterjee, S. Sachdev, and M. S. Scheurer, arXiv: 1812.08792 (2018).
\bibitem{Xu} S. Chatterjee, H. Guo, S. Sachdev, R. Samajdar, M. S. Scheurer, N. Seiberg and C. Xu, arXiv:1903.01992 (2019).
\bibitem{ZXLi} Z. X. Li and D. H. Lee , arXiv:1905.04248 (2019).
\bibitem{LCO}R. Coldea et al., Phys. Rev. Lett. \textbf{86}, 5377 (2001).
\bibitem{Fermion}Here we adopt a Bosonic RVB picture rather than the Fermionic RVB picture. In the Fermionic RVB picture, the spinon would develop a large SDW gap in the presence of the static antiferromagnetic order and would be inconsistent with the monotonic increase of the thermal Hall signal at low temperature.
\bibitem{Messio}L. Messio, C. Lhuillier, and G. Misguich, Phys. Rev. B \textbf{87}, 125127 (2013).
\bibitem{LDA}S. D. Liang, B. Doucot and P. W. Anderson, Phys. Rev. Lett. \textbf{61}  365(1988).
\bibitem{Li1}Tao Li, to appear.
\end{thebibliography}
\end{document}